\documentclass{article}

\usepackage{amsmath} 
\usepackage{amssymb} 
\usepackage[mathcal,mathscr]{eucal} 
\usepackage{eufrak} 
\usepackage{epsfig} 
\usepackage{amsthm} 
\usepackage{graphicx} 
\usepackage[numbers, sort&compress]{natbib} 
 
\input{tcilatex}

\begin{document}

\vspace{10 mm}

\begin{center}
{\large {\bf The Pioneer Anomaly: The Measure of a Topological Phase Defect of Light in Cosmology}}

\bigskip

{\large J.L. Rosales (JoseLuis.Rosales@xerox.com)}
 {\footnote{ 
I dedicate this paper to my daughter Ana }}

\bigskip

{\large Quantum Information Group, Spanish Royal Society of Physics}
\end{center}

\vspace{3mm}

\begin{center}
{\bf Abstract}
\end{center}

\begin{quote}
It is shown that a wave vector representing a light pulse in an
adiabatically evolving expanding space should develop, after a round trip
(back and forth to the emitter) a geometric phase for helicity states at a
given fixed position coordinate of this expanding space.In a section of the
Hopf fibration of the Poincar\'{e} sphere{\bf \ }${\bf S}^{2}$ that
identifies a projection to the physically allowed states, the evolution
defines a parallel transported state\ that can be joined continuously with
the initial state by means of\ the associated Berry-Pancharatnam connection.
The connection allows to compute an anomaly in the frequency for the vector
modes in terms of the scale factor of the space-time background being
identical to the reported Pioneer Anomaly.
\end{quote}

\bigskip

\vspace{2mm} {\it {\bf {\ \ 1.Introduction}}}

Analysis of the radio-metric tracking data from the Pioneer 10/11 spacecraft
at distances between 20-70 astronomical units (AU) from the Sun has
consistently indicated the presence of an anomalous, small, constant Doppler
frequency drift. The drift is a blue-shift, uniformly changing with rate $\
a_{t}=(2.92\pm \ 0.44)\times 10^{-18}s/s2$.\ It can also be interpreted as a
constant acceleration of $\ \ a_{P}=(8.74\pm \ 1.33)\times 10^{-8}cm/s2$ \
apparently directed towards the Sun \cite{kn:Anderson 1998}, \cite%
{kn:Anderson 2002}. This signal has become known as the Pioneer Anomaly
since it does not seem to correspond to standard \ Newtonian dynamics (as
far as this kind of anomalous acceleration has never been found perturbing
the orbits of the planets in the Solar System.)

There were attempts to explain the anomaly based on the recently discovered
accelerated expansion of the Universe. This association was motivated by the
numerical coincidence that links the magnitude of the Pioneer Anomaly to the
product of the Hubble constant and the speed of light; on the other hand,
since Hubble's flow would be for the probes vanishing small in comparison
with their typical velocities in the scale of the Solar System, the effect
can not be originated from Hubble's dynamics acting {\it on the probes}
{\footnote{$\delta \omega/\omega \simeq - h (R/c)(v_P/c)^2$, $h$ 
denoting the local expansion rate, $R/c$, standing for the Doppler delay of 
lihgt signals from the spacecraft at $R$ and $v_P$ the probe velocity. See the Appendix.}.
Moreover,{\it \ intuitively}, such a mechanism would produce an opposite
sign for the effect. Yet, let us look at the problem again disregarding the
motion of particles in Hubble's flow, i.e., from a geometrical perspective,
upon studying the internal states of light themselves during a measurement
of the spacecraft position. Need our intuition still be valid?

In order to escape from such common intuitions, this paper analyses the true
effect \ derived from the existence of some small, non vanishing, local
space expansion rate on the adiabatic evolution of internal states of light.
We will not be concerned here on the exact physical meaning of such an
hypothesis, rather, we want to obtain the expected measurable consequences
of it. Very surprisingly a proof will be given below \ in the sense that 
{\it such an effect does exist} and that {\it a locally expanding space time
originates a blue shift frequency anomaly of geometrical origin in the phase
of light} being the same for every possible polarization state.The result is
in full agreement with some earlier heuristic\ proposals \cite%
{kn:Rosales-SanchezGomez} \cite{kn:Rosales 2004}. Consequently
theoretical and future experimental work remains to be made in order to
clarify the consequences of the present proof and the measurement of the
anomaly.

Let us start, then, upon considering an expanding space-time with metric
given by{\footnote{ notice that the metric (1) might be the special case, for 
$\chi(t)=e^{c(\Lambda/3)^{1/2}t}$ and
 $M/R \ll 1$, of
the Schwarzschild-DeSitter metric
\begin{eqnarray*}
ds^2=(1-2m/r-\frac{\Lambda}{3} r^2)c^2d\tau^2-\frac{1}{1-2m/r-\frac{\Lambda}{3} r^2}dr^2
 \end{eqnarray*}
for $r=R\chi(t)$, $\tau= t-\frac{1}{2 c(\Lambda/3)^{1/2}} \ln(1-\frac{\Lambda}{3}R^2\chi(t)^2)$
and $M(t)=m\chi(t)$,
that is
\begin{eqnarray*}
ds^2\simeq (1-2M(t)/R)dt^2-\frac{1}{1-2M(t)/R}\chi^2(t)dR^2
\end{eqnarray*}
} %
\begin{equation}
ds^{2}=c^{2}dt^{2}-\chi (t)^{2}(dR^{2}+R^{2}d\Omega ^{2})%
\mbox{\hspace{
2 mm}.}
\end{equation}%
We are interested on the mathematical description of the measurable phase of
a beam of light immersed in this space-time at a given constant coordinate
position $R$, after a round trip of total time $T\hspace{0in}\hspace{0in}$%
has taken place. This corresponds to the measurement of the approximate
distance $cT/2$ to some remote mirror on which light bounces at $t=T/2$,
back to the emitter as in the tracking signal experiment to the space
probes. The problem is, then, different from that of obtaining the phase of
a light wave emitted from a distant source (galaxies, say.) In such an
expanding space it is well known that there exists a red shift in the
frequency of such distant light source which is Hubble's law. Now, instead,
we want to compare the phase of a given photon emitted from $R$ \ at time $%
t=0\hspace{0in}$ $\hspace{0in}$and observed at this same position after a
round trip of time $t=T$. In order to cope with this problem, we should
model $\chi (t)$ as a external slow adiabatic parameter for the internal
evolution of the phase states of light. This is obviously a geometric
(instead of dynamic) problem for the internal state of light that can be
visualized as that of determining the parallel transported state of light
after such an adiabatic evolution in the external parameter space has taken
place.

\bigskip

{\bf 2. Helicity states of light: the Hopf fibration of }${\bf S}^{2}$%
\bigskip

In order to enlighten the solution to this problem, we need a completely new
mathematical framework. What we require now \ is a description of light in
terms of its internal helicity complex vector state ${\frak h}$, and we
intend to separate the dynamic (temporal) fast evolution from the
geometrical (adiabatic) evolution of the internal phase state of light.

The polarization complex vector satisfy

\begin{equation}
{\frak h}{=(}x,y)\ ,{\frak h}^{\dagger }\cdot {\frak h}{\bf =\ }1%
\mbox{\hspace{ 2
mm}.}
\end{equation}%
\ $x,y$ being complex numbers. It might also conveniently be described in
terms of the two component spinor 
\begin{equation}
|\Psi >=\QATOPD[ ] {\Psi _{+}}{\Psi _{-}}\mbox{\hspace{ 2 mm},}
\end{equation}%
for, $\Psi _{\pm }\equiv \frac{1}{\sqrt{2}}(x\pm iy)\exp (i\beta )$. Thus $%
<\Psi |\Psi >=1$, and $\beta $ an arbitrary phase. Each such $|\Psi >$ is
the eigenvector of some Hermitian matrix (the polarization matrix
"Hamiltonian") of the form

\begin{equation}
{\bf r\cdot \sigma =}\left( 
\begin{array}{cc}
z & x-iy \\ 
x+iy & -z%
\end{array}%
\right) =\left( 
\begin{array}{cc}
\cos (\theta ) & \sin (\theta )\exp (-i\phi ) \\ 
\sin (\theta )\exp (+i\phi ) & -\cos (\theta )%
\end{array}%
\right) \mbox{\hspace{ 2
mm}.}
\end{equation}%
where $\sigma ${\bf \ }is the vector of Pauli matrices and $\ {\bf r=(x,y,z)}
$ \ is a unit vector with polar angles $\theta $ and $\phi $. The
geometrical coordinates $(\theta ,\phi )$ define the Poincar\'{e} sphere $\ 
{\bf S}^{2}$. The relevant spinor is an eigenstate of the Hermitian matrix
above times an arbitrary phase,

\begin{equation}
|\Psi (t)>=\QATOPD[ ] {\cos (\theta /2)}{\sin (\theta /2)e^{i\phi }}\exp
(i\beta )\mbox{\hspace{ 2 mm},}
\end{equation}%
Which is a vector in some enlarged space that will be defined below. To this
new space belongs a set of complex spinors with three degree of freedom
(hereafter we follow \cite{kn:Aitchison 1987})

\begin{equation}
|\Psi >=\QATOPD[ ] {x_{1}+ix_{2}}{x_{3}+ix_{4}}\mbox{\hspace{ 2 mm},}
\end{equation}%
such that

\begin{equation}
x_{1}^{2}+x_{2}^{2}+x_{3}^{2}+x_{4}^{2}=1\mbox{\hspace{ 2 mm},}
\end{equation}%
the $x_{i\text{ }}$ are rather the coordinates of a point on an ${\bf S}%
^{3}. $ Upon defining $\psi =\beta +\phi ,$ we get from (5), (6) and (7) 
\begin{eqnarray*}
x_{1} &=&\cos (\theta /2)\cos (\beta ) \\
x_{2} &=&\cos (\theta /2)\sin (\beta ) \\
x_{3} &=&\sin (\theta /2)\cos (\psi ) \\
x_{4} &=&\sin (\theta /2)\sin (\psi )\mbox{\hspace{ 2 mm},}
\end{eqnarray*}%
The metric of this space is

\begin{equation}
ds^{2}=\frac{1}{4}d\theta ^{2}+\cos ^{2}(\theta /2)d\beta ^{2}+\sin
^{2}(\theta /2)d\psi ^{2}\mbox{\hspace{ 2 mm},}
\end{equation}%
thus, we have enlarged the two polarization degree of freedom to three by
means of a ${\it U(1)}$ gauge field $\beta .\hspace{0in}$ $\hspace{0in}{\bf S%
}^{3}$ \ can be regarded as a principal bundle with base space ${\bf S}^{2}$
and a ${\it U(1)\hspace{0in}\hspace{0in}}$ structure group. This procedure
is called Hopf fibration of ${\bf S}^{2}.$

On the other hand, in order to identify these geometrical state coordinates
with relevant physical quantities defined in the physical space-time, recall
that, for the metric (1), the Eikonal of a light wave at the space-time
point of physical coordinates $(R,T)$ is given by{\footnote{%
It is easily verified. Given that, 
\begin{eqnarray*}
\Xi &=&-g_{\mu\nu}k^{\mu}x^{\nu}
\end{eqnarray*}%
and that, for $k_0=\omega/c$ 
\begin{eqnarray*}
k_{\mu}k^{\mu}=\omega^2/c^2-\chi^2 (k^R)^2=0 ,
\end{eqnarray*}%
it follows that
\begin{eqnarray*}
(k^R)^2=(\omega/c\chi)^2 \\
\end{eqnarray*}
and
\begin{eqnarray*}
k_R=\pm \chi\omega/c
\end{eqnarray*}
} }

\begin{equation}
\Xi (r,t)=-[\omega T-\frac{\omega }{c}{\bf l\cdot R}\chi ]%
\mbox{\hspace{2
mm}.}
\end{equation}%
Where ${\bf l\cdot R}=\pm R$ \ for circular positive and negative
polarization states corresponding to the North and the South Pole of ${\bf S}%
^{2}$. In the spinor formalism, it is equivalent to taking $\beta =-\phi
(R,T)/2$,\hspace{0.01in} $\psi =+\phi (R,T)/2$ and $\phi (R,T)=-2\frac{%
\omega }{c}\chi R$. \ Restricting $\beta $ and $\psi \hspace{0in}$ in this
way is called taking a section ${\frak C}\hspace{0in}\hspace{0.01in}$ \ of
the fibre bundle. Therefore, after (9)

\begin{equation}
|\Psi _{+}>=|\epsilon _{+}>\exp (+i\frac{\omega }{c}R\chi ),
\end{equation}%
and

\begin{equation}
|\Psi _{-}>=|\epsilon _{-}>\exp (-i\frac{\omega }{c}R\chi ),
\end{equation}%
($|\epsilon _{+}>\equiv \QATOPD[ ] {1}{0}$ and $|\epsilon _{-}>\equiv \QATOPD%
[ ] {0}{1}$.) These are the counterpropagating (positive and negative
helicities) wave modes as required for an electromagnetic field propagating
in \ a space-time whose metric is given by \ equation.(1).

In order to clarify the formalism let us introduce the "magnetic-like field" 
{\footnote {
Hereafter we will use the notation
$\frak \Delta$-variation with respect to the external parameter $\chi$, 
it should correspond to a geometrical variation. 
this is different with respect to $d$-differenciation.}}

\begin{equation}
d{\frak B}=\frak \Delta\hspace{0in}[\phi /2]
\end{equation}%
then, we rewrite equations. (10) and (11) as

\begin{equation}
|\Psi (\chi )>=e^{-i\int d\hspace{0in}{\frak S}}|\Psi >\mbox{\hspace{ 2 mm},}
\end{equation}%
for the "Hamiltonian"%
\begin{equation}
d{\frak S}\equiv d{\frak B}\cdot \sigma _{3}\mbox{\hspace{ 2 mm},}
\end{equation}%
$|\Psi (\chi )>$ is then, the solution of a Schr\"{o}dinger equation%
\begin{equation}
id|\Psi (\chi )>=d{\frak S}|\Psi (\chi )>\mbox{\hspace{ 2 mm}.}
\end{equation}

Equations (10)-(15) define the unitary evolution of the polarization state
in the expanding space-time. Notice that the evolution of the helicity
states of light, satisfying Maxwell equations, within an expanding
space-time is formally equivalent to that of a quantum spinor. We will now
explore more deeply this similarity.

\bigskip

\vspace{2mm} {\it {\bf {\ \ 3.Adiabatic evolution: the Berry Connection.}}}

Let us consider changes in the state of polarization accomplished
continuously and consider the curved space-time as a dielectric medium. If
the dielectric variation $\chi (t)\hspace{0in}\hspace{0in}$ is slow enough,
the beam remains in a polarization state. In a cycle on the Poincar\'{e}
sphere the displacement unit vector ${\frak h}$ \ \ will be accompanied by a
phase, the Pancharatnam's phase given by

\begin{equation}
<A|A^{\prime }>=\exp [-i\frac{\Omega (C)}{2}]\mbox{\hspace{ 2 mm}.}
\end{equation}%
where, the cycle $C$ connects the state $|A>$ with$\hspace{0in}|A^{\prime }>$
and $\Omega (C)$ is the solid angle of the circuit $C$ on the Poincar\'{e}
sphere. During adiabatic evolution, the local eigenstates are continued by
means of the differential equation

\begin{equation}
{\frak h}^{\dagger }\cdot d\hspace{0in}{\frak h}{\bf =}0%
\mbox{\hspace{ 2
mm},}
\end{equation}%
i.e., the polarization state is parallel transported through the cycle. This
is\ consequence of the field being governed \ by Maxwell's equations as was
shown by Berry\cite{kn:Berry 1987}. Using the definitions of the spinor Sch%
\"{o}dinger-like evolution of the previous section, we see that, (after
subtraction of the trivial dynamic contribution) one equivalently obtains 
\begin{equation}
<\overset{\symbol{126}}{\Psi }|d\overset{\symbol{126}}{\Psi }>{\bf =\hspace{%
0in}}0\mbox{\hspace{ 2
mm},}
\end{equation}%
for 
\begin{equation}
{\bf |}\overset{\symbol{126}}{{\bf \Psi }}{\bf (T)}{\bf >\equiv }\exp
\{i\tint\nolimits_{0}^{T}\overset{\wedge }{{\frak H}}({\bf t}^{\prime })d%
{\bf t}^{\prime }\}{\bf |\Psi (T)>}\mbox{\hspace{ 2
mm},}
\end{equation}%
and

\[
\overset{\wedge }{{\frak H}}[t^{\prime }]{\bf =\frac{\partial }{\partial t^{\prime }}\func{Re}<\Psi
(\chi ,t^{\prime }}{\bf )|\{}\int_{1}^{\chi (t^{\prime })}{\bf d{\frak S}(}%
\chi )\}{\bf |\Psi (\chi ,t^{\prime }})>{\bf .} 
\]%
Equation (18) is the Berry connection{\footnote{%
Clairly,${\bf |}\overset{\symbol{126}}{{\bf \Psi }}{\bf (t)}{\bf >}$
satisfies the equation%
\[
id{\bf |}\overset{\symbol{126}}{{\bf \Psi }}{\bf (t)}{\bf >=\{d}{\frak S}%
{\bf {\frak -}}\overset{\wedge }{{\frak H}}(t)dt\}{\bf |\overset{\symbol{126}%
}{\Psi }(t)>.} 
\]%
Contracting this with ${\bf <\overset{\symbol{126}}{\Psi }(t)|}$ (also using 
${\bf <\overset{\symbol{126}}{\Psi }}|\overset{\symbol{126}}{\Psi }>=1$)
yields the paralell transport law (18).}}. We are now interested on the
adiabatic evolution of a spinor state when $\beta (\chi )$ $\ $and $\psi
(\chi )\hspace{0in}$ vary continuously and slowly enough so that if the
system is initially in the state ${\bf |\Psi >}$, eigenstate of the
Hamiltonian, ${\bf |\Psi (\chi )>}$ will also instantaneously be an
eigenstate of the same Hamiltonian; this is the condition of the adiabatic
theorem. The formal equivalence of this system with the quantum spinor
allows using Berry's theorem\cite{kn:Berry 1984}: During adiabatic
evolution, the total phase change of ${\bf |\Psi (\chi )>}$ round ${\frak %
C(\beta ,\psi )}$ is given by

\begin{equation}
|\Psi (\chi )>=\exp (i\gamma ({\frak C}))\exp [-i\int_{1}^{\chi }d{\frak S}%
(\chi )]|\Psi ({\bf 1})>\mbox{\hspace{ 2 mm},}
\end{equation}%
for $\gamma ({\frak C})$, the Berry phase given by{\footnote{%
\[
{\bf |}\overset{\symbol{126}}{{\bf \Psi }}{\bf (1)}{\bf >=\exp \{-i[}\int 
{\bf d}{\frak S}{\bf {\frak -}}\overset{\wedge }{{\frak H}}(t)dt]\}{\bf %
|\Psi (1)>.} 
\]%
}}

\begin{equation}
\gamma ({\frak C})=i\int_{{\frak C}}<\Psi |\frak \Delta\hspace{0in}\Psi >%
\mbox{\hspace{
2 mm},}
\end{equation}%
\bigskip 
\begin{equation}
{\bf |}\overset{\symbol{126}}{{\bf \Psi }}{\bf (\chi )>=}\exp (i\gamma (%
{\frak C})){\bf |}\overset{\symbol{126}}{{\bf \Psi }}{\bf (1)>}%
\mbox{\hspace{ 2
mm},}
\end{equation}%
or, after Equation (5),%
\begin{equation}
\gamma ({\frak C})=-\int_{{\frak C}}\cos ^{2}(\theta /2)\frak \Delta\beta +\sin
^{2}(\theta /2)\frak \Delta\psi \mbox{\hspace{ 2 mm}.}
\end{equation}%
This defines a vector potential%
\begin{equation}
\gamma ({\frak C})=-\int_{{\frak C}}{\bf A\cdot \frak \Delta r}
\end{equation}%
where, via the ${\bf S}^{3}$ metric in Equation (8)

\begin{equation}
A_{\theta }=0,\hspace{0in}A_{\beta }=\cos (\theta /2),\hspace{0in}A_{\psi
}=\sin (\theta /2).
\end{equation}%
These potentials are manifestly non-singular. For helicity states, $\theta
=0,$ and $\theta =\pi $%
\begin{equation}
\gamma ({\frak C})=-\int_{{\frak C}}\frak \Delta\beta 
\mbox{ \hspace{ 2 mm},\hspace{2
mm}}\theta =0
\end{equation}%
\begin{equation}
\gamma ({\frak C})=-\int_{{\frak C}}\frak \Delta\psi 
\mbox{ \hspace{ 2 mm},\hspace{2
mm}}\theta =\pi
\end{equation}%
for the section ${\frak C}$ of the fibre bundle $\beta (\chi )=-$ $\psi
(\chi )$, equivalently $\beta =-\phi /2$, i.e., the physically allowed
states, this obtains%
\begin{equation}
\gamma ({\frak C})=\pm \frac{{\frak \Delta }\phi (\chi )}{2}\simeq \mp R%
\frac{\omega }{c}\overset{\cdot }{\chi }T\mbox{ \hspace{ 2 mm},}
\end{equation}%
Recall ${\frak \Delta }\phi =-2R\frac{\omega }{c}[\chi (T)-1]$. This phase
shift corresponds to positive, negative helicities respectively. Moreover,
for\ a general polarization state we get%
\begin{equation}
\gamma ({\frak C})=\cos (\theta )\frac{{\frak \Delta }\phi }{2}=-\cos
(\theta )R\frac{\omega }{c}\overset{\cdot }{\chi }T\mbox{ \hspace{ 2 mm}.}
\end{equation}

\bigskip

\bigskip

{\bf 4}{\it {\bf {.The frequency Anomaly.}}}

\bigskip

We will now obtain a remarkable consequence of equations (22) and (29). The
parallel transported positive and negative helicity states are given by 
\begin{equation}
{\bf |}\overset{\symbol{126}}{{\bf \Psi }}{\bf (T)>=}\exp [\mp iR\frac{%
\omega }{c}(\overset{\cdot }{\pm \chi }T-1)]|\varepsilon _{\pm }>.
\end{equation}%
and, as a result of the Berry connection,%
\begin{equation}
<\overset{\symbol{126}}{\Psi }|d\overset{\symbol{126}}{\Psi }>{\bf =}0%
\mbox{\hspace{ 2
mm},}
\end{equation}%
one directly gets, for both cases 
\begin{equation}
\overset{\cdot }{\omega }_{\pm }=\omega _{\pm }\overset{\cdot }{\chi }
\end{equation}%
For general polarization states we get again from \ equations (22) and (29)%
\begin{equation}
{\bf |}\overset{\symbol{126}}{{\bf \Psi }}{\bf (\theta ,T)>=}e^{-i\cos
(\theta )R\frac{\omega }{c}\overset{\cdot }{\chi }T}\{\cos (\theta
/2)e^{iR\omega /c}|\varepsilon _{+}>+\sin (\theta /2)e^{-iR\omega
/c}|\varepsilon _{-}>\}.
\end{equation}%
And the Berry connection (31) indicates that%
\begin{equation}
\cos ^{2}(\theta /2)a_{\chi }(\theta ,T)+\sin ^{2}(\theta /2)b_{\chi
}(\theta ,T)=0
\end{equation}%
where $a_{\chi }(\theta ,T)=-\frac{\overset{\cdot }{\omega }}{\omega }(1-%
\overset{\cdot }{\chi }T\cos \theta )+\overset{\cdot }{\chi }\cos (\theta )%
\hspace{0in}\hspace{0in}$ and $\hspace{0in}b_{\chi }(\theta ,T)=\frac{%
\overset{\cdot }{\omega }}{\omega }(1+\overset{\cdot }{\chi }T\cos \theta )+%
\overset{\cdot }{\chi }\cos (\theta ).$ \ Solving this for $\frac{\overset{%
\cdot }{\omega }}{\omega }$ we finally obtain the general expression of the
frequency anomaly%
\begin{equation}
\overset{\cdot }{\omega }_{\theta }=\omega _{\theta }\overset{\cdot }{\chi }%
+O(\overset{\cdot }{\chi }^{2}),
\end{equation}%
which is independent of $\hspace{0in}\theta \hspace{0in}$ for every
polarization state. This remarkable result coincides with the observed blue
shift known as the Pioneer anomaly for $a_{t}=\overset{\cdot }{\chi }.$

\bigskip
\newpage

{\bf 5. Conclusions.}

The proof given here indicates that the "Pioneer effect" detected in radar
signals \cite{kn:Anderson 1998}, \cite{kn:Anderson 2002}, \cite{kn:Anderson
2004b} should have nothing to do with the probe but only with the fact that
the spacecraft is acting as a "mirror" for light signals, thus, being \ a
consequence of the adiabatic evolution of the internal states of light
locally monitored by the global expanding space-time metric. The Anomaly
should more properly be described as corresponding to the measure of a
topological phase defect of light. There is a formal identification between
the polarization vector temporal evolution in an expanding space with that
of a two state quantum spinor, so we explore their common algebra to
discover that, upon considering the expanding space as an adiabatic
dielectric, it yields to obtaining a "quantum Berry's phase" in a section of
the Hopf fibration of the Poincar\'{e} sphere{\bf \ }${\bf S}^{2}$ that
identifies a projection to the physically allowed states. It becomes the
frequency blue shift anomaly upon using the algebra of parallel transported
states (i.e., the Berry connection). Thus, the Doppler anomalous phase shift
finds its explanation on the grounds of a Berry phase, a geometrical effect.
This is just a non dynamic element of the evolution of helicity states of
light in the expanding space background. This demonstrates entirely that the
effect should not affect to the planets but only to light and that it is
wrongly interpreted as a dynamic acceleration, being fully equivalent to a
calibration effect similarly to the Foucault Pendulum angle defect in
measuring Earth rotation (the Hannay's angle which, indeed, is the classical
analog to the Berry's quantum phase, also a geometrical effect). In this
sense, light rays play a similar r\^{o}le in the expanding space than
Foucault's Pendulum does while determining Earth's rotation. On the other
hand, given that the result has nothing to do with dynamics, it does not
violate Birkhoff's theorem. Moreover, the anomaly only depends on
the "Time of Flight" of light , since the location of the spacecraft has not
enter into the proof. This predicts that a geostationary system of
satellites (LISA mission, for instance) or perhaps other specific more
advanced mission as recently proposed would obtain the same result 
\cite{kn:Anderson 2004a},\cite{kn:Anderson 2004b}. A completely new type of
optical experiments becomes, indeed, also possible, for instance, optical
laser ranging with active mirrors able to accumulate this phase topological
defect in n-way round trips configurations. A kind of such an experimental
arrangement could be a future sophistication of the Moon Laser Ranging
Device. 

\begin{center}
{\bf ACKNOWLEDGMENTS}
\end{center}
On the occation of the 70th birthday of Professor Santos, it is a pleasure to contribute with this paper to this volume.
I want to give special thanks to professors Antonio Fernandez-Ra\~{n}ada and Slava Turyshev from
stimulating discussions. I thank specially to Antonio Fernandez-Ra\~{n}ada
for telling me about Aitchison's work. I am also very thankful to Xerox Corporation.

\begin{center}
{\bf APPENDIX}
\end{center}
We want to obtain in this appendix the order of magnitude of the perturbing effect, for the  motion of a probe
with radial velocity 
$v_P$, of  the existence of a local expanding space-time. In order to do this, let us scale out the expansion of the space for the metric 
$ds^2=c^2dt^2-\chi(t)^2dR^2$, i.e., let us consider $R_{*}\equiv \chi R$.  The physical meaning of this choice of 
coordinates is that, during the motion of the spacecrafts, we use the Newtonian metric as parametrically static. In terms of this scaled
radial coordinate, one gets
\begin{eqnarray*}
ds^2=c^2(1-R_{*}^2/c^2 h^2)dt^2-dR_{*}^2+2h R_{*}dR_{*}dt
\end{eqnarray*} 
and $h\equiv \dot{\chi}/\chi$. This defines the radial vector 
$g_{*}\equiv -g_{0R_{*}}/g_{*00}\simeq -hR_{*}/c$.
 
The radial velocity of the probe is, using the scaled coordinates,
\begin{eqnarray*}
(g_{*00})^{1/2}v_{*}=\frac{dR_{*}}{d\tau}
\end{eqnarray*}
where $d\tau=dt-g_{*}dR_{*}/c$ is the proper time at $R_{*}$ for this curved space-time.
This obtains
\begin{eqnarray*}
\dot{R}_{*}\simeq v_{*}(1+hR_{*}v_{*}/c^2) +O(\dot{\chi}^2).
\end{eqnarray*}
The Doppler expected effect for the probes is
\begin{eqnarray*}
\omega^{'}=\omega (1-\dot{R}_{*}/c)\simeq\omega (1-v_{*}/c)-\omega h(R_{*}/c)(v_{*}/c)^2.
\end{eqnarray*}
This corresponds to an anomalous red shift $\delta \omega/\omega \simeq -h t(v_{*}/c)^2$ for $t=R_{*}/c$. 
That is why  the Pioneer Anomaly can not be originated from the dynamic effect of the expansion acting on the probes. 
The link between the figures of Hubble's ($h$) and Anderson's ($a_t$) constants can not be dynamic.

\begin{thebibliography}{99}
\bibitem{kn:Anderson 1998} Anderson, J.D., Laing, P.A., Lau, E.L.,Liu, A.S.,
Nieto,M.M., and Turyshev,S.G., {\em Phys. Rev. Lett.} {\bf 81}, 2858,
(1998), e-print gr-qc/9808081.

\bibitem{kn:Anderson 2002} Anderson, J.D., Laing, P.A., Lau, E.L.,Liu, A.S.,
Nieto,M.M., and Turyshev,S.G., {\em Phys. Rev.} {\bf D65},082004/1-50
(2002), e-print gr-qc/0104064.

\bibitem{kn:Rosales-SanchezGomez} Rosales, J.L.and S\'anchez-G\'omez, J.L.
e-print gr-qc/9810085.

\bibitem{kn:Rosales 2004} Rosales, J.L. Proceedings of the International
Conference on the Pioneer Anomaly, ZARM, University of Bremen, May 2004.
e-print gr-qc/0401014.

\bibitem{kn:Aitchison 1987} Aitchison,I.J.R. {\em Acta Phys. Pol.}{\bf B}18,
207, (1987).

\bibitem{kn:Berry 1987} Berry,M. V. {\em J. Mod. Opt.}{\bf 34},11, 1401
(1987).

\bibitem{kn:Berry 1984} Berry, M. V. {\em Proc. R. Soc.} {\bf A392}, 45
(1984).

\bibitem{kn:Anderson 2004a} H. Dittus, C. Lammerzahl, S. Theil, B. Dachwald,
W. Seboldt, W. Ertmer, E. Rasel, R. Foerstner, U. Johann, F. W. Hehl, C.
Kiefer, H.-J. Blome, R. Bingham, B. Kent, T. J. Sumner, O. Bertolami, J. L.
Rosales, B. Christophe, B. Foulon, P. Touboul, P. Bouyer, S. Reynaud, C. J.
de Matos, C. Erd, J. C. Grenouilleau, D. Izzo, A. Rathke, J. D. Anderson, S.
W. Asmar, S. G. Turyshev, M. M. Nieto, and B. Mashhoon, "A Consolidated
Cosmic Vision Theme Proposal to Explore the Pioneer Anomaly," submitted to
ESA (Oct. 2004).

\bibitem{kn:Anderson 2004b} Anderson, J.D., Nieto, M.M., and Turyshev,S.G.,
e-print:gr-qc/0411077
\end{thebibliography}
\end{document}